\documentclass[
    aps,
    physrev,
    a4paper,
    twocolumn,
    superscriptaddress,
    amsmath,amssymb
    ]{revtex4-2}

\usepackage{graphicx} 
\usepackage{subfig}
\usepackage{geometry}
\usepackage[utf8]{inputenc}
\usepackage[T1]{fontenc}
\usepackage{setspace}
\usepackage{amsmath}
\usepackage{bbm}
\usepackage{float}
\usepackage{microtype}
\usepackage{amssymb}
\usepackage{amsthm}

\usepackage{tikz}
\usepackage{mathtools}
\usepackage{enumerate}
\usepackage{hyperref}
\usepackage{xcolor}

\newtheorem{theorem}{Theorem}

\newcommand{\denop}{\mathcal{D}}
\newcommand{\hs}{\mathcal{H}}
\newcommand{\linop}{\mathcal{L}}
\DeclareMathOperator{\mytr}{Tr} 

\newcommand{\sumkl}{\sum_{k,l = 0}^{d-1}}
\newcommand{\sumj}{\sum_{j = 0}^{d-1}}
\newcommand{\Pkl}{P_{k,l}}
\newcommand{\ckl}{c_{k,l}}
\newcommand{\Wkl}{W_{k,l}}
\newcommand{\id}{\mathbbm{1}}
\newcommand{\ZZ}{\mathbbm{Z}}
\newcommand{\FF}{\mathbbm{F}}

\newcommand{\PPT}{\mathrm{PPT}}
\newcommand{\NPT}{\mathrm{NPT}}
\newcommand{\SEP}{\mathrm{SEP}}
\newcommand{\ENT}{\mathrm{ENT}}

\begin{document}

\title{Noise-robust 1-copy distillation protocol for all distillable Bell-diagonal qutrits}

\author{Tobias C. Sutter}
\email{tobias.christoph.sutter@univie.ac.at}
\affiliation{University of Vienna, Faculty of Physics, Währingerstrasse 17, 1090 Vienna, Austria.}
\author{Christopher Popp}%
\email{christopher.popp@univie.ac.at}
\affiliation{University of Vienna, Faculty of Physics, Währingerstrasse 17, 1090 Vienna, Austria.}
\author{Beatrix C. Hiesmayr}
\email{beatrix.hiesmayr@univie.ac.at}
\affiliation{University of Vienna, Faculty of Physics, Währingerstrasse 17, 1090 Vienna, Austria.}
\affiliation{IT:U Interdisciplinary Transformation University, Freistädter Strasse 400, 4040 Linz, Austria.}

\date{\today}

\begin{abstract}
Entanglement distillation is the process of converting noisy entangled states into maximally entangled pure states via local operations and classical communication.
A long-standing, unresolved question is which entangled states are amenable to distillation, known as the distillability problem.
We solve this for Bell-diagonal qutrits with Weyl structure, and present a noise-robust scheme for entanglement distillation.
In particular, we find that violating the positive partial transposition (PPT) criterion is necessary and sufficient for the 1-distillability of these states.
For this, we construct a Schmidt rank 2 eigenvector of the partially transposed density matrix associated with its unique, three-fold degenerate negative eigenvalue.
This feature makes the derived entanglement distillation protocol resilient to white-noise effects on the quantum states.
Our results thus make noisy entangled qutrit pairs more accessible for future quantum technologies.
\end{abstract}

\maketitle

\section{Introduction}

Entanglement is a fundamental resource for many quantum technologies, such as quantum computation \cite{jozsa_role_2003} and cryptography \cite{zapatero_advances_2023}.
The performance of these technologies relies heavily on the amount of entanglement available \cite{vidal_efficient_2003}, and for bipartite applications, the gold standard is typically maximally entangled pure states.
However, experimental noise inevitably degrades the entanglement of any such pure states, yielding mixed states that are often less useful in practice, as the attainable quantum advantage may be lost \cite{preskill_quantum_2018}.

One practical solution to this problem is entanglement distillation, i.e., using local operations and classical communication (LOCC) to convert multiple copies of some input state into fewer copies of maximally entangled states.
The difficulty for entanglement distillation in practice and theory is twofold.
First, for the distillation task, one requires at least partial knowledge of the given quantum state \cite{zang_no-go_2025}.
Thus, to design highly performant distillation protocols, suitable noise models leading to those expected mixed resource states need to be studied.
One important model concerns generalized Pauli errors (via the probabilistic application of Weyl-Heisenberg matrices) acting on one subsystem of the maximally entangled state, yielding Bell-diagonal states.
Second, to know whether some state can be distilled at all requires checking that, for some number $n$ of copies, one can obtain a maximally entangled state by LOCC.
Ref.~\cite{watrous_many_2004} shows that for every such $n>0$, there exists a state $\rho$ that is $n$-copy-undistillable, but $(n+1)$-copy-distillable, illustrating one of the difficulties of the distillability question.

Adding to this difficulty is the fact that not all entangled states can be distilled.
A necessary criterion for distillability is the violation of the positive partial transposition (PPT) criterion \cite{horodecki_separability_1996, peres_separability_1996}.
The problem is that while all separable states satisfy the criterion, not all entangled states violate it for bipartite systems with dimension $d_A \times d_B$ with $d_A d_B >6$.
This means that such states are entangled but undistillable, and are called bound entangled \cite{horodecki_mixed-state_1998} (for a recent review on bipartite bound entanglement, see \cite{hiesmayr_bipartite_2025}).
Nonetheless, for bipartite qubit-qudit systems with $d_A \geq 2$ and $d_B=2$, a violation of the PPT criterion is necessary and sufficient for distillability \cite{dur_distillability_2000}.
A long-standing open problem in this regard is whether a violation of the PPT criterion is also sufficient for distillability in general dimensions (with evidence against it presented, e.g., in Refs.~\cite{dur_distillability_2000, divincenzo_evidence_2000}).

In this Letter, we answer this question in the affirmative for Bell-diagonal qutrit states with Weyl structure and present a corresponding noise-robust construction that can be used for entanglement distillation.
In particular, we explicitly construct a 1-distillability witness vector for all Bell-diagonal qutrit-qutrit states $\rho$ that violate the PPT criterion.
This is based on an eigenvector of the partial transpose of $\rho$ corresponding to its unique, three-fold degenerate negative eigenvalue.
It has the advantage that detecting the 1-distillability of $\rho$ via the associated distillability witness and implementing the derived local projections as an initial step for entanglement distillation are more robust against white noise in real experiments than other approaches.

\section{Entanglement distillation, partial transpose, and entanglement witnesses \label{sec:ent_dist}}

Consider a bipartite Hilbert space $\hs = \hs_A \otimes \hs_B$ with $d_A :=\dim(\hs_A)$ and $d_B:=\dim(\hs_B)$.
We denote linear operators on $\hs$ by $\linop(\hs)$, and the set of density matrices by $\denop(\hs) = \{\rho \in\linop(\hs) \,|\, \rho \geq 0, \mytr(\rho)=1\}$.
A state $\rho\in\denop(\hs)$ is called \textit{separable} if it can be written as $\rho = \sum_{i} p_i \,\rho_i^{(A)} \otimes \rho_i^{(B)}$ with $\rho_i^{(A)} \in \denop(\hs_{A})$ and $\rho_i^{(B)} \in \denop(\hs_{B})$, and \textit{entangled} otherwise.
We denote the sets of separable and entangled states by $\SEP$ and $\ENT$, respectively.
If unambiguous, we abbreviate product vectors such as $|\psi_A\rangle \otimes |\psi_B\rangle$ by $|\psi_A, \psi_B\rangle$.

Entanglement distillation is the task of converting $n$ copies of a resource state $\rho\in\denop(\hs)$ into $m\leq n$ copies of a maximally entangled target state $\Tilde{\rho}$ using only local quantum operations and classical communication (LOCC).
If this is possible for some $n,m>0$, we say that $\rho$ is \emph{distillable}.
Without loss of generality, one can choose the maximally entangled qubit state $\Tilde{\rho} = |\phi^+\rangle\langle \phi^+|$, where $|\phi^+\rangle = \frac{1}{\sqrt{2}} \sum_{i=0}^{1} |i,i\rangle$ \cite{hiesmayr_bipartite_2025}.

A valuable tool for investigating the distillability of quantum states is the partial transposition, which we denote by $\rho^\Gamma = (\id_A \otimes T_B)(\rho)$, where $T_B: |i\rangle\langle j| \mapsto |j\rangle\langle i|$ is the transposition on $\linop(\hs_B)$ in the computational basis.
We define the disjoint sets $\PPT:=\{\rho\in\denop(\hs) \,|\, \rho^\Gamma \geq 0\}$ and $\NPT:=\{\rho\in\denop(\hs) \,|\, \rho^\Gamma \ngeq 0\}$ corresponding to states having a positive and a non-positive partial transposition, respectively.
This provides a partition of the set of density matrices $\denop(\hs) = \PPT \cup \NPT$.

A necessary condition for $\rho\in\denop(\hs)$ to be distillable is that $\rho\in\NPT$ \cite{horodecki_mixed-state_1998}, which is also sufficient if $\min(d_A, d_B) =2$ \cite{dur_distillability_2000}.
Beyond that, the following gives an exact characterization of distillability for general dimensions:
\begin{theorem} [\cite{horodecki_mixed-state_1998,dur_distillability_2000}] \label{thm:n-distillability}
    A bipartite state $\rho\in\denop(\hs)$ is distillable if and only if there exists $n>0$ and $|\varphi\rangle \in \hs^{\otimes n}$ with Schmidt rank 2 such that
    \begin{align} \label{eq:n-distillability}
        \langle\varphi | \; (\rho^\Gamma)^{\otimes n} \; |\varphi\rangle <0 \;.
    \end{align}
    W.l.o.g., $|\varphi\rangle$ can be chosen to be normalized, in which case its Schmidt decomposition is
    \begin{align} \label{eq:dist_witness_vec}
        |\varphi\rangle = \mu_0 \, |a_0\rangle \otimes |b_0\rangle + \mu_1 \, |a_1\rangle\otimes |b_1\rangle    
    \end{align}
    with $\mu_0 \geq \mu_1 >0$, $\mu_0^2 + \mu_1^2 =1$, $\{|a_0\rangle , |a_1\rangle\} \subset \hs_A^{\otimes n}$, and $\{|b_0\rangle , |b_1\rangle\} \subset \hs_B^{\otimes n}$ satisfying $\langle a_i |a_j\rangle = \langle b_i | b_j\rangle = \delta_{i,j}$.
\end{theorem}
We call a state \emph{$n$-distillable} if $n$ is the smallest positive integer such that Theorem~\ref{thm:n-distillability} holds, and the accompanying Schmidt rank 2 vector $|\varphi\rangle$ is its \emph{$n$-distillability witness vector}.
If $\rho$ is 1-distillable, we can rewrite \eqref{eq:n-distillability} as
\begin{gather}\label{eq:dist-witness_eq}
    \begin{aligned}
        0>\langle\varphi|\rho^\Gamma |\varphi\rangle &= \mytr(\rho^\Gamma |\varphi\rangle\langle\varphi|) \\
    &= \mytr(\rho \,(|\varphi\rangle\langle\varphi|)^\Gamma) = \mytr(\rho \, W_\varphi) \;,
    \end{aligned}
\end{gather}
where we defined the \emph{1-distillability witness}
\begin{gather}\label{eq:dist-witness_def}
    \begin{aligned}
        W_\varphi &:= (|\varphi\rangle\langle\varphi|)^\Gamma \\
        &= \mu_0^2 \, |a_0, b_0^\ast\rangle \langle a_0, b_0^\ast| + \mu_1^2 \, |a_1, b_1^\ast\rangle \langle a_1, b_1^\ast| \\
        & \hspace{5mm}+ \mu_0 \, \mu_1 (|a_0, b_1^\ast\rangle \langle a_1, b_0^\ast| + |a_1, b_0^\ast\rangle \langle a_0, b_1^\ast|) \;.
    \end{aligned}
\end{gather}
Here, $^\ast$ denotes entry-wise complex conjugation in the computational basis.
From this, we see that $\mathrm{spec}(W_\varphi) = \{\mu_0^2, \mu_1^2, \mu_0 \mu_1, -\mu_0 \mu_1,0\}$, and hence $W_\varphi\ngeq0$.
Furthermore, for all product vectors $|a,b\rangle:=|a\rangle \otimes |b\rangle \in \hs$, we have
\begin{gather}
    \begin{aligned}
            \mytr( |a,b\rangle \langle a,b| W_\varphi) &= \mytr( |a,b^\ast \rangle\langle a,b^\ast |\varphi\rangle\langle \varphi|)  \\
            &= |\langle a,b^\ast| \varphi\rangle |^2 \geq 0 \;,
    \end{aligned}
\end{gather}
implying that $W_\varphi$ is an entanglement witness.
This operator is also decomposable \cite{lewenstein_optimization_2000} and thus cannot detect PPT-entangled states.
Moreover, it is weakly optimal in the sense that there exist product vectors $|a,b\rangle \in \hs$ such that $\langle a,b|W_\varphi |a,b\rangle = 0$.
In particular, this is achieved by $|a\rangle = |a_0\rangle$ and $|b\rangle = |b_1^\ast\rangle$.
Finally, the mirrored operator to $W_\varphi$ is $M_\varphi = \mu_0^2 \, \id_{d^2} - W_\varphi$ \cite{bera_structure_2023}, which is positive semidefinite and therefore not an entanglement witness.

For qutrit-qutrit systems, there is a sufficient condition for 1-distillability: 
\begin{theorem}[\cite{chen_distillability_2016}] \label{thm:1-dist_qutrit-qutrit}
    If $\rho\in\NPT$ with $d=d_A=d_B=3$ and $\rho^\Gamma$ has at least two non-positive eigenvalues counting multiplicities, then $\rho$ is 1-distillable.
\end{theorem}
The proof of this theorem in Ref.~\cite{chen_distillability_2016} invokes Theorem~\ref{thm:n-distillability} and constructs a Schmidt rank 2 vector $|\varphi\rangle$ such that $\langle\varphi|\rho^\Gamma|\varphi\rangle <0$.
However, it is not concerned with minimizing $\langle\varphi|\rho^\Gamma|\varphi\rangle$ over suitable $|\varphi\rangle$.
While this is sufficient for proving Theorem~\ref{thm:1-dist_qutrit-qutrit}, it hinders its experimental applicability in noisy conditions, as we discuss next.

\subsection{Experimental limitations of Theorem~\ref{thm:1-dist_qutrit-qutrit} \label{sec:experimental_limitations_of_theorem_2}}

Since experimental realizations of qutrits are inevitably affected by noise, the main practical limitation for using $|\varphi\rangle$ from Ref.~\cite{chen_distillability_2016} stems from the fact that $\langle\varphi|\rho^\Gamma|\varphi\rangle<0$ might be close to zero.
Below, we illustrate this for (a) the verification of 1-distillability using the witness $W_\varphi$ and (b) entanglement distillation using local projections derived from $|\varphi\rangle$, and argue that for both tasks it may be beneficial to minimize $\langle\varphi|\rho^\Gamma|\varphi\rangle$.

For (a), modeling experimental imperfections for $\rho$ as white noise, we define
\begin{align}
    \rho_\mathrm{noisy} := (1-p) \, \rho + \frac{p}{d^2}\,\id_{d^2} \;,
\end{align}
with noise parameter $p\in[0,1]$.
The expectation value of the 1-distillability witness \eqref{eq:dist-witness_def} for this state is
\begin{align}
    \mytr\left( W_\varphi \rho_\mathrm{noisy} \right) = (1-p) \, \langle \varphi| \rho^\Gamma |\varphi\rangle + \frac{p}{d^2} \;.
\end{align}
The state $\rho_\mathrm{noisy}$ is detected as 1-distillable whenever $\mytr\left( W_\varphi \, \rho_\mathrm{noisy} \right) <0$, or equivalently, when
\begin{align} \label{eq:tolerable_noise}
    p < p_{\rho, \max} = \frac{-d^2 \, \langle \varphi| \rho^\Gamma |\varphi\rangle}{1-d^2 \,\langle \varphi| \rho^\Gamma |\varphi\rangle} \;,
\end{align}
where we used $d\geq 2$ and $-\frac{1}{2} \leq\lambda_{\min}(\rho^\Gamma) \leq \langle \varphi| \rho^\Gamma |\varphi\rangle<0$ for $\rho\in\NPT$ \cite{rana_negative_2013}.
As the right-hand side of \eqref{eq:tolerable_noise} is monotonically decreasing for $-\frac{1}{2}\leq\langle \varphi| \rho^\Gamma |\varphi\rangle<0$, the tolerable noise level $p$ for witnessing 1-distillability of $\rho_\mathrm{noisy}$ is maximized when $\langle \varphi| \rho^\Gamma |\varphi\rangle$ is minimized.

Concerning (b), we can construct local rank-2 projection operators
\begin{align}
    P_A = \sum_{i=0}^{1} |a_i \rangle\langle a_i| \;, \quad
    P_B = \sum_{i=0}^{1} |b_i^\ast \rangle\langle b_i^\ast | \;,
\end{align}
from the 1-distillability witness vector $|\varphi\rangle$ in \eqref{eq:dist_witness_vec}.
These can be used to filter the original state $\rho$ to an effective entangled NPT qubit-qubit state
\begin{align} \label{eq:sigma}
    \sigma:= q^{-1} (P_A \otimes P_B) \,\rho\, (P_A \otimes P_B) \;,
\end{align}
where $q = \mytr((P_A \otimes P_B) \,\rho)$ is the success probability of the filtering process.
The minimal eigenvalue of $\sigma^\Gamma$ is bounded from above by
\begin{align} \label{eq:min_eval_sigma}
    \lambda_{\min} (\sigma^\Gamma)
    \leq \langle\varphi| \sigma^\Gamma |\varphi\rangle
    = q^{-1} \langle\varphi|\rho^\Gamma|\varphi\rangle <0 \;,
\end{align}
where we used $(P_A \otimes P_B^T) |\varphi\rangle = |\varphi\rangle$.
Because all entangled bipartite qubit states can be distilled, this filtering can be used as the initial step of an entanglement distillation protocol for $\rho$.
However, experimental noise can degrade the entanglement of $\sigma$.
If we model this decoherence process as white noise on $\sigma$, we retain $\sigma\in\NPT$ for higher noise levels if $\lambda_{\min}(\sigma^\Gamma)$ is as negative as possible.
Hence, in light of \eqref{eq:min_eval_sigma}, minimizing $\langle\varphi|\rho^\Gamma|\varphi\rangle$ over Schmidt rank 2 vectors $|\varphi\rangle$ may be crucial for designing practical entanglement distillation protocols that function reliably in the presence of experimental noise.

In this regard, the minimum of $\langle\varphi|\rho^\Gamma|\varphi\rangle$ for any $|\varphi\rangle$ is lower bounded by the smallest eigenvalue $\lambda_{\min}(\rho^\Gamma)$.
If this extreme value is reached, $|\varphi\rangle$ is an eigenvector of $\rho^\Gamma$.
If, furthermore, $|\varphi\rangle$ is a 1-distillability witness vector, the inequality in \eqref{eq:min_eval_sigma} becomes an equality because $|\varphi\rangle$ is also an eigenvector of $\sigma^\Gamma$ corresponding to its unique negative eigenvalue $\lambda_{\min}(\sigma^\Gamma) = q^{-1} \lambda_{\min}(\rho^\Gamma)$.
In this case, assuming white noise on $\sigma$, the noise threshold for maintaining $\sigma\in\NPT$ is even larger than the one in \eqref{eq:tolerable_noise} if $q<4/d^2$.
To see this, we consider white noise affecting $\sigma$ (viewed as a bipartite qubit state), and define
\begin{align}
    \sigma_\mathrm{noisy} := (1-p) \, \sigma + \frac{p}{4} \, \id_{4} \;,
\end{align}
with $p\in[0,1]$.
We have $\sigma_\mathrm{noisy} \in \NPT$ if and only if $\lambda_{\min}(\sigma_\mathrm{noisy}^\Gamma) <0$, or equivalently
\begin{align}
    p < p_{\sigma, \max} =\frac{-4 \, \lambda_{\min}(\sigma^\Gamma)}{1-4 \,\lambda_{\min}(\sigma^\Gamma)} = \frac{-4 \, \lambda_{\min}(\rho^\Gamma)}{q-4 \,\lambda_{\min}(\rho^\Gamma)} \;,
\end{align}
where we used $\lambda_{\min}(\sigma^\Gamma) = q^{-1} \lambda_{\min}(\rho^\Gamma)$.
For the qubit state $\sigma$ to be more noise-resilient than $\rho$, we require $p_{\sigma, \max} > p_{\rho, \max}$, which is the case if and only if $q < 4/d^2$, as claimed.

\section{Bell-diagonal states with Weyl structure} \label{sec:BDS}

The mixed states of interest in this contribution are Bell-diagonal states that inherit group structure via the projective Weyl-Heisenberg group \cite{sutter_group-theoretic_2026}.
They occur naturally by considering generalized Pauli errors.
The $d$-dimensional Weyl-Heisenberg operators via which they are defined are given by
\begin{align}
    \Wkl &= \sumj \omega^{jk} \, |j\rangle\langle j+l| \;, \label{eq:Wkl}
\end{align}
where $\omega = e^{2\pi i/d}$.
Throughout the paper, we denote the ring of integers modulo $d$ by $\ZZ_d := \ZZ / d\ZZ$, and we take all indices and computational basis vector entries modulo $d$.
The operators \eqref{eq:Wkl} satisfy the so-called Weyl relations
\begin{gather}
    \begin{aligned} \label{eq:Weyl_relations}
        W_{i,j} W_{k,l} &= \omega^{jk} \, W_{i+k,j+l} \;,\\
        \Wkl^\ast &= W_{-k,l} \;, \\
        \Wkl^T &= \omega^{-kl} W_{k,-l} \;.
    \end{aligned}
\end{gather}
By defining the canonical maximally entangled state as
\begin{align}
    |\Omega_{0,0}\rangle = \frac{1}{\sqrt{d}} \sum_{i=0}^{d-1} |i,i\rangle \;, \label{eq:Omega_00}
\end{align}
we construct a basis of $\hs = \hs_A \otimes \hs_B$ with $d_A = d_B = d$ using the $d^2$ maximally entangled states
\begin{align} \label{eq:Omega_kl}
    |\Omega_{k,l}\rangle = (W_{k,l}\otimes \id) |\Omega_{0,0}\rangle \;,
\end{align}
where $k,l\in \ZZ_d$.
Convex combinations of the projectors $\Pkl := |\Omega_{k,l}\rangle\langle\Omega_{k,l}|$ are the so-called Bell-diagonal states, forming a ``magic simplex'' \cite{baumgartner_special_2007}, given by
\begin{align}
    \mathcal{M}_d &= \left\{ \rho = \sumkl \ckl \, \Pkl \; \Big| \; \sumkl \ckl = 1, \; \ckl \geq 0 \right\} \;. \label{eq:magic_simplex}
\end{align}
Operationally, states in $\mathcal{M}_d$ can be obtained by locally applying the quantum channel $\mathcal{E}(X)=\sum_{k,l=0}^{d-1} c_{k,l} \, W_{k,l}\, X\, W_{k,l}^\dagger$ to the maximally entangled state $P_{0,0}$, i.e., $(\mathcal{E} \otimes \id_d)(P_{0,0}) \in \mathcal{M}_d$.
Any $\rho\in\mathcal{M}_d$ is diagonalized by the Bell-unitary
\begin{align}\label{eq:U_Bell_unitary}
    U = \sum_{r,s=0}^{d-1} |r,s\rangle\langle\Omega_{r,s}| \;. 
\end{align}
A straightforward computation reveals a block-diagonal structure of the partially-transposed state
\begin{align} \label{eq:rho_gamma_block_diagonal}
    \rho^\Gamma = U^\dagger \left(\sum_{m=0}^{d-1} |m\rangle\langle m| \otimes B_m \right) U \;,
\end{align}
where we defined
\begin{gather}\label{eq:B_m}
    \begin{aligned}
        B_m &:= \frac{1}{d^2} \sum_{k,l,x,y=0}^{d-1} \omega^{y(k-m)-xl} c_{k,l+y} W_{x,2y}  \\
        &= \frac{1}{d} \sum_{k,l,y=0}^{d-1} \omega^{y(k-m)} c_{k,l} |l-y\rangle\langle l+y| \;.
    \end{aligned}
\end{gather}
For odd dimension $d$, the hermitian blocks $B_m$ are unitarily equivalent; for even $d$, there are two classes of unitarily equivalent blocks, one for even and one for odd $m$ \cite{baumgartner_special_2007}.
Using \eqref{eq:B_m} and \eqref{eq:Weyl_relations}, it can be checked that for all $d$ we have
\begin{align} \label{eq:Bm->Bm+2}
    B_{m+2} = W_{1,0} B_m W_{1,0}^\dagger \;.
\end{align}
As shown in Ref.~\cite{sutter_group-theoretic_2026}, any Bell-diagonal NPT state $\rho\in\mathcal{M}_3 \cap \NPT$ has three negative eigenvalues and hence satisfies the premise of Theorem~\ref{thm:1-dist_qutrit-qutrit}, leading to the corollary that every $\rho\in\mathcal{M}_3 \cap \NPT$ is 1-distillable, i.e., there exist no NPT bound entangled states in $\mathcal{M}_3$.
In the next section, we give a constructive proof of our main theorem, which is the stronger statement
\begin{theorem} \label{thm:main_thm}
    Let $\rho\in\mathcal{M}_3 \cap \NPT$ and let $\lambda_{\min}(\rho^\Gamma)<0$ be the smallest eigenvalue of $\rho^\Gamma$.
    Then there exists a normalized Schmidt rank 2 vector $|\varphi\rangle$ such that
    \begin{align}
        \rho^\Gamma |\varphi\rangle = \lambda_{\min}(\rho^\Gamma) |\varphi\rangle \;,
    \end{align}
    and hence
    \begin{align}
        \langle\varphi|\rho^\Gamma|\varphi\rangle = \lambda_{\min}(\rho^\Gamma) <0 \;.
    \end{align}
    Therefore, every $\rho\in\mathcal{M}_3 \cap \NPT$ is 1-distillable.
\end{theorem}


\section{Proof of Theorem~\ref{thm:main_thm} \label{sec:construction}}

To prove Thm.~\ref{thm:main_thm}, we start by noting that $\rho\in\mathcal{M}_3 \cap \NPT$ implies $\rho^\Gamma \ngeq 0$, and hence $B_m \ngeq 0$ for some $m\in\ZZ_3$ by \eqref{eq:rho_gamma_block_diagonal}.
As $d=3$ is odd, the $B_m$ are unitarily equivalent, and thus $B_m \ngeq 0$ for all $m\in \ZZ_3$.
Let $|u_m\rangle$ be the normalized eigenvector of $B_m$ with the smallest eigenvalue, i.e., $B_m |u_m\rangle = \lambda_{\min}(\rho^\Gamma) |u_m\rangle$.
Using \eqref{eq:Bm->Bm+2}, these eigenvectors satisfy
\begin{align} \label{eq:relation_u_m+2_with_u_m}
    |u_{m+2}\rangle = W_{1,0} |u_m\rangle \;.
\end{align}
Next, we consider the normalized vector
\begin{align}\label{eq:psi}
    |\psi\rangle = \sum_{i=0}^{2} \psi_i \; |i\rangle \otimes |u_i\rangle \;,
\end{align}
satisfying $\sum_{i=0}^{2} |\psi_i|^2 = 1$ and
\begin{gather}
    \begin{aligned}
        \langle \psi |U\rho^\Gamma U^\dagger |\psi\rangle
        &=  \sum_{m=0}^{2}
        |\psi_m|^2
        \underbrace{\langle u_m | B_m |u_m\rangle}_{=\lambda_{\min}(\rho^\Gamma)}\\
        &= \lambda_{\min}(\rho^\Gamma) <0 \;. \label{eq:dist_witness_eq_0}
    \end{aligned}
\end{gather}
Next, we rewrite \eqref{eq:U_Bell_unitary} as
\begin{align}
    U = (F \otimes \id) \, C_s \;,
\end{align}
with the discrete Fourier transform $F$ and the controlled-sum gate $C_s$ defined as
\begin{align}
    F &:= \frac{1}{\sqrt{3}} \sum_{x,y=0}^{2} \omega^{-xy} |x\rangle\langle y| \;, \\
    C_s &:= \sum_{z=0}^{2} |z\rangle\langle z| \otimes W_{0,z} \;.
\end{align}
Note that $C_s (|i, j\rangle) = |i, j-i\rangle$.
Furthermore, we define
\begin{align}
    |\alpha^{(m)}\rangle := F^\dagger |u_m\rangle \;.
\end{align}
Using \eqref{eq:relation_u_m+2_with_u_m} and $F^\dagger W_{k,l} F = \omega^{-kl} W_{-l,k}$ we find
\begin{gather}
    \begin{aligned}\label{eq:beta_and_alpha_connection}
    |\alpha^{(m+2)}\rangle &= F^\dagger |u_{m+2}\rangle = F^\dagger W_{1,0} |u_m\rangle \\
    &= F^\dagger W_{1,0} F |\alpha^{(m)}\rangle = W_{0,1} |\alpha^{(m)}\rangle \;.
    \end{aligned}
\end{gather}
Thus, writing $\alpha^{(m)}_k = \langle k|\alpha^{(m)}\rangle$, we have $\alpha^{(m+2)}_i = \alpha^{(m)}_{i+1}$.
Now we make the ansatz
\begin{align}
    |\Tilde{\varphi}\rangle = \sum_{i,k=0}^{2} \, \psi_i \, \alpha^{(i)}_k \, |k,k+i\rangle \;.
\end{align}
A direct computation yields $U |\Tilde{\varphi}\rangle = \FF |\psi\rangle$, with the flip operator $\FF = \sum_{r,s=0}^{2} |r,s\rangle\langle s,r|$.
Hence, defining
\begin{align}
    |\varphi\rangle := U^\dagger \FF U |\Tilde{\varphi}\rangle \;,
\end{align}
we obtain with \eqref{eq:dist_witness_eq_0}
\begin{align}
    0 > \langle\psi| \, U \, \rho^\Gamma \, U^\dagger \, |\psi\rangle = \langle \varphi |\rho^\Gamma |\varphi\rangle \;.
\end{align}
By construction, this $|\varphi\rangle$ is a normalized eigenvector of $\rho^\Gamma\ngeq0$ with negative eigenvalue $\lambda_{\min}(\rho^\Gamma)<0$, i.e., $\rho^\Gamma |\varphi\rangle = \lambda_{\min}(\rho^\Gamma) |\varphi\rangle$.
To show that it is also a 1-distillability witness vector for $\rho\in\mathcal{M}_3\cap \NPT$, and thereby prove Thm.~\ref{thm:main_thm}, we have to show that $|\Tilde{\varphi}\rangle$ has Schmidt rank 2 for some choice of $\psi_i$ in \eqref{eq:psi}, and that $U^\dagger \FF U$ does not change the Schmidt rank.
First, we rewrite $U^\dagger \FF U$ as
\begin{align}
    U^\dagger \FF U = (F^\dagger \otimes F) \FF \;,
\end{align}
which implies that it does not change the Schmidt rank. 
Second, to show that $|\Tilde{\varphi}\rangle$ has Schmidt rank 2, we write $|\Tilde{\varphi}\rangle = \sum_{i,j=0}^{2} C_{i,j} |i,j\rangle$ and investigate its coefficient matrix.
It suffices to only consider the case $\psi_0=1/\sqrt{2}, \psi_1=0,\psi_2=-1/\sqrt{2}$, for which we obtain
\begin{align} \label{eq:C-matrix}
    C = \frac{1}{\sqrt{2}}\begin{pmatrix}
        \alpha_0^{(0)} & 0 & -\alpha_0^{(2)} \\
        -\alpha_1^{(2)} & \alpha_1^{(0)} & 0 \\
        0 & -\alpha_2^{(2)} & \alpha_2^{(0)}
        \end{pmatrix} \;.
\end{align}
The vector $|\Tilde{\varphi}\rangle$ has Schmidt rank 2 if and only if $C$ has rank 2, which is the case if and only if $\det(C) =0$ and at least one $2\times 2$ (principal) minor is nonzero \cite{horn_matrix_2012}.
The quantities of interest are thus
\begin{align}
    \det(C) &= \frac{1}{2\sqrt{2}}\left(\alpha_0^{(0)} \alpha_1^{(0)} \alpha_2^{(0)} - \alpha_0^{(1)}\alpha_1^{(1)}\alpha_2^{(1)}\right), \label{eq:detC}\\
    M_0 &= \frac{1}{2}\det\begin{pmatrix}
        \alpha_1^{(0)} & 0 \\
        -\alpha_2^{(2)} & \alpha_2^{(0)}
        \end{pmatrix} = \frac{1}{2} \left(\alpha_1^{(0)} \alpha_2^{(0)}\right) , \label{eq:minor0}\\
    M_1 &= \frac{1}{2}\det\begin{pmatrix}
        \alpha_0^{(0)} & -\alpha_0^{(2)} \\
        0 & \alpha_2^{(0)}
        \end{pmatrix} = \frac{1}{2}\left(\alpha_0^{(0)} \alpha_2^{(0)}\right) , \label{eq:minor1}\\
    M_2 &= \frac{1}{2} \det\begin{pmatrix}
        \alpha_0^{(0)} & 0 \\
        -\alpha_1^{(2)} & \alpha_1^{(0)}
        \end{pmatrix} = \frac{1}{2}\left( \alpha_0^{(0)} \alpha_1^{(0)}\right) , \label{eq:minor2}
\end{align}
where $M_j$ for $j\in\ZZ_3$ are $2\times 2$ principal minors of $C$.
Using $\alpha^{(i+2)}_k = \alpha^{(i)}_{k+1}$ from \eqref{eq:beta_and_alpha_connection}, we find that $\det(C) = 0$ and thus $\mathrm{rank}(C)\leq 2$.
Furthermore, at least one $M_j$ does not vanish, which can be seen as follows: $|\alpha^{(m)}\rangle = F^\dagger |u_m\rangle$ is an eigenvector of $F^\dagger B_m F$ with negative eigenvalue $\langle \alpha^{(m)} | F^\dagger B_m F |\alpha^{(m)}\rangle = \lambda_{\min}(\rho^\Gamma) <0$.
Computing $F^\dagger B_m F$, we find that it has non-negative diagonal entries
\begin{align}
    \langle k| F^\dagger B_m F |k\rangle = \frac{1}{3} \sum_{l=0}^{2} c_{m-k,l} \geq 0 \;.
\end{align}
Hence, $|\alpha^{(m)}\rangle$ cannot be proportional to a computational basis state, and thus at least two $\alpha_i^{(m)}$ must be non-zero, leading to at least one $M_j\neq0$ and $\mathrm{rank}(C)\geq2$.
Consequently, $\mathrm{rank}(C)=2$ and we found a Schmidt rank 2 vector $|\varphi\rangle = U^\dagger \FF U |\Tilde{\varphi}\rangle$ such that $\langle \varphi | \rho^\Gamma |\varphi\rangle = \lambda_{\min}(\rho^\Gamma) <0$.
This proves Thm.~\ref{thm:main_thm}.


\section{Conclusion and outlook \label{sec:conclusion}}

We showed that all Bell-diagonal qutrit-qutrit states with non-positive partial transposition are 1-distillable, and explicitly constructed a Schmidt rank 2 vector $|\varphi\rangle$ that serves as a 1-distillability witness vector for such states.
Consequently, there is no NPT bound entanglement for this subclass of states.
The crucial feature of our construction is that $|\varphi\rangle$ is an eigenvector of the partially-transposed state $\rho^\Gamma$ associated with its unique, three-fold degenerate negative eigenvalue.
Hence, it minimizes the expectation value of the 1-distillability witness $W_\varphi = |\varphi\rangle\langle\varphi|^\Gamma$, allowing for noise-resilient 1-distillability certification.
Furthermore, it enables projecting the original resource state onto an entangled bipartite qubit state whose negativity equals the negative eigenvalue of $\rho^\Gamma$ divided by the success probability of the local filters constructed from $|\varphi\rangle$.
With this initial step, an effective entanglement distillation protocol is achieved.
In particular, the resulting qubit pair can be locally transformed to an NPT qubit Werner state \cite{clarisse_entanglement_2006}, which is LOCC-equivalent to an NPT qubit isotropic state.
Subsequently applying the stabilizer-based two-copy distillation protocol FIMAX \cite{popp_novel_2025}, which is numerically shown to be able to distill all such states \cite{popp_low-fidelity_2025}, provides a concrete route to maximally entangled qubit states.

In this regard, we note that directly applying FIMAX to an NPT Bell-diagonal qutrit pair can fail, even though the state is 1-distillable.
Indeed, Ref.~\cite{popp_low-fidelity_2025} shows that FIMAX can distill only 83\% of low-fidelity NPT Bell-diagonal qutrits, whereas the above-described strategy succeeds for all such states.
This is noteworthy insofar as FIMAX exploits the same group structure as is present in Bell-diagonal systems and is therefore tailor-made for this state family, for which it shows superior performance compared to other recurrent two-copy protocols.
Our results thus reveal the limitations of this stabilizer-based distillation protocol, leaving open the question of whether an algorithmic distillation protocol that works for all $n$-distillable states exists, or whether, in some cases, projecting onto an entangled qubit pair must be considered as an initial step.
Furthermore, it would be interesting to investigate the properties that distinguish FIMAX-distillable from FIMAX-undistillable states, as this could help develop a distillation protocol that applies to all Bell-diagonal states.

Regarding the limitations of the present results, the most striking one is the dimensionality: While our construction works for all $\rho\in\mathcal{M}_3 \cap \NPT$, it generally fails for $d>3$ as the Schmidt rank of the constructed $|\varphi\rangle$ is $d-1$ (with slightly different $\psi_i$ than above if $d$ is even).
Even though this is not directly useful for distilling the $d$-dimensional Bell-diagonal state $\rho\in\mathcal{M}_d \cap \NPT$, it can nonetheless be used as an initial step to LOCC-transform any NPT Bell-diagonal qudit to a $(d-1)$-dimensional NPT Werner state.

Finally, our results further strengthen the evidence that Bell-diagonal states are highly useful resource states for entanglement distillation compared with other state families, such as Werner states.
In this sense, Werner states appear comparatively hard to distill; indeed, they are often regarded as extremal in this respect, since any NPT state can be transformed by LOCC into an NPT Werner state.
Bell-diagonal states, by contrast, seem to be substantially more amenable to distillation.
Since they also arise naturally in experimental settings under generalized Pauli noise, our findings further support their relevance as practically accessible resource states for entanglement distillation.


\begin{acknowledgments}
This research was funded in whole or in part by the Austrian Science Fund (FWF)[10.55776/COE1, 10.55776/P36102].
For open access purposes, the author has applied a CC BY public copyright license to any author accepted manuscript version arising from this submission.
\end{acknowledgments}


\bibliography{references}

\end{document}